\documentclass[preprint,showpacs,preprintnumbers,amsmath,amssymb,superscriptaddress]{revtex4}

\usepackage{graphicx}% Include figure files
\usepackage{dcolumn}% Align table columns on decimal point
\usepackage{amsmath}    % need for subequations
\usepackage{verbatim}   % useful for program listings
\usepackage{color}      % use if color is used in text
\usepackage{subfigure}  % use for side-by-side figures
\usepackage{hyperref}   % use for hypertext links, including those to external documents and URLs
\usepackage{bm}% bold math

\begin{document}

%\preprint{Toshiba Research Europe Limited/Confidential}

\title{Interference of dissimilar photon sources
}% Force line breaks with \\

\author{A. J. Bennett}
\email{anthony.bennett@crl.toshiba.co.uk}
\affiliation{Toshiba Research Europe Limited, Cambridge Research Laboratory,\\
208 Science Park, Milton Road, Cambridge, CB4
OGZ, U. K.}

\author{R. B. Patel}
\affiliation{Toshiba Research Europe Limited, Cambridge Research Laboratory,\\
208 Science Park, Milton Road, Cambridge, CB4
OGZ, U. K.}
\affiliation{Cavendish Laboratory, Cambridge University,\\
JJ Thomson Avenue, Cambridge, CB3 0HE, U. K.}

\author{C. A. Nicoll}
\affiliation{Cavendish Laboratory, Cambridge University,\\
JJ Thomson Avenue, Cambridge, CB3 0HE, U. K.}

\author{D. A. Ritchie}
\affiliation{Cavendish Laboratory, Cambridge University,\\
JJ Thomson Avenue, Cambridge, CB3 0HE, U. K.}

\author{A. J. Shields}
\affiliation{Toshiba Research Europe Limited, Cambridge Research Laboratory,\\
208 Science Park, Milton Road, Cambridge, CB4
OGZ, U. K.}

\date{\today}%

\begin{abstract}
If identical photons meet at a semi-transparent
mirror they appear to leave in the same
direction, an effect called ``two-photon
interference". It has been known for some time
that this effect should occur for photons
generated by dissimilar sources with no common
history, provided the measurement cannot
distinguish between the photons\cite{Mandel83}.
Here we report a technique to observe such
interference with isolated, unsynchronized
sources whose coherence times differ by several
orders of magnitude. In an experiment we
interfere photons generated via different
physical processes, with different photon
statistics. One source is stimulated emission
from a tuneable laser, which has Poissonian
statistics and a $neV$ bandwidth. The other is
spontaneous emission from a quantum dot in a
p-i-n diode \cite{Yuan02, Bennett05} with a $\mu
eV$ linewidth. We develop a theory to explain the
visibility of interference, which is primarily
limited by the timing resolution of our
detectors.
\end{abstract}


\maketitle %this tag has to go at end of title section

Two-photon interference is at the heart of many
optical quantum information processing protocols
but to be scalable these proposals require large
numbers of identical photons in pre-determined
states. This leads to the question: what is the
minimum requirement to observe interference
between two sources? Since the first
demonstration of two photon
interference\cite{Hong87} the proto-typical
experimental arrangement has used parametric
down-conversion to create photon pairs by
exciting a crystal with a strong laser.
Experiments have been reported where a heralded
photon interferes with a weak Poissonian source
\cite{Rarity97, W07}, though in these cases both
photon streams were derived from the same laser.
Recently, interference between distinct but
nominally identical down-conversion sources has
been demonstrated with separate pump lasers
\cite{Halder07, Kaltenbaek06}. On-demand
zero-dimensional photon sources are attractive
for scalable quantum-information processing as
they naturally create only one photon at a time
\cite{Lounis05}, however care must be taken to
ensure spectral jitter and dephasing do not
distinguish each photon. Indistinguishable
single-photon emission was demonstrated with a
semiconductor quantum dot \cite{Santori02} and a
single atom \cite{Legero04} by optically exciting
a single source twice in quick succession. More
recently, it has been shown that two identical
zero-dimensional sources excited with the same
laser can generate indistinguishable photons
\cite{Beugnon06, Maunz07, Chaneliere07,
Sanaka09}. From a fundamental point of view it
would be interesting to be able to interfere
sources that are not similar. Such interference
effects can be used to determine the spectral
density matrix of a single photon \cite{W07}.
Also potential applications of quantum
information processing will in future require
interactions between distributed sources, either
for quantum computing \cite{Lim05} or interfacing
weak lasers commonly used in quantum cryptography
with the more exotic sources being investigated
for quantum repeaters and memories. Here we
report a demonstration of interference between
non-classical emission from an electrically
excited InAs quantum dot and a commercially
available semiconductor laser.

\begin{figure}
\includegraphics[width = 140mm]{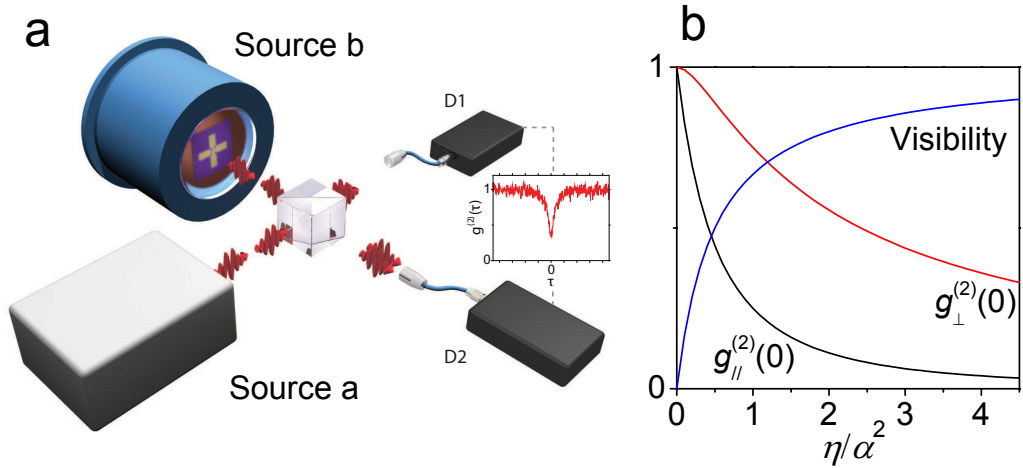} % Here is how to import EPS art
\caption{\label{Fig1} Interference of a
single-photon source with a weak Poissonian
source. (a) Shows a schematic of the experiment.
(b) Shows the predicted correlation function at
zero time delay, $g^{(2)}(0)$ for parallel
(black) and orthogonal (red) photons as a
function of the intensity ratio of the sources,
$\eta/\alpha^{2}$. Shown in blue is the resulting
visibility of two-photon interference,
$(g^{(2)}_{\bot}(0)-g^{(2)}_{\|}(0))/g^{(2)}_{\bot}(0)$.}
\end{figure}

To understand our experiment it is useful to
consider idealized sources incident from opposite
sides on a 50/50 semi-transparent mirror (as
illustrated in Fig. \ref{Fig1} (a)). Two
detectors at the outputs of the mirror measure
the probability of two photons leaving the mirror
at the same time. If we assume photons from both
sources have perfect overlap in energy, space and
time but no mutual coherence then we must only
consider the probabilities of the sources
emitting a certain number of photons. The laser
obeys Poissonian statistics,
$\langle\psi_{a}|\psi_{a}\rangle =
\exp\left(-\alpha^{2}\right)\sum_{n_{a}}\frac{\alpha^{2n_{a}}}{n_{a}!}\langle
n_{a}|n_{a}\rangle$ and for the quantum light
source $\langle\psi_{b}|\psi_{b}\rangle =
\left(1-\eta\right)\langle 0_{b}|0_{b}\rangle +
\eta\langle 1_{b}|1_{b}\rangle$. In our notation
$\eta$ and $\alpha^{2}$ are proportional to the
probabilities of detecting either a photon from
the anti-bunched source or the laser,
respectively, at the output of the experiment.
For simplicity we only expand the coherent state
up to $n_{a}=2$, which is valid for a strongly
attenuated laser $\alpha^{2}\ll1$. If the sources
do not interfere there are a number of ways the
detectors can collect one photon each: either
they collect both photons from the laser (with
probability $2RT\alpha^{4}/4$) or they collect
one photon from the laser and one from the single
photon source. This second possibility can occur
if both photons are reflected (with probability
proportional to $R^{2}\eta\alpha^{2}/2$) or if
they are both transmitted
($T^{2}\eta\alpha^{2}/2$). However, if the two
sources are indistinguishable then interference
will occur leading the latter two terms to cancel
when $R=T$. These joint-detection probabilities
$g^{(2)}(0)$ are normalized to the probability of
detecting two photons at different times, $(\eta
+ \alpha^{2})^{2}$. Equation \ref{equation1}
gives the probability of a detecting a photon in
both outputs at the same time when the sources
have parallel polarisations and are
indistinguishable. A useful control measurement
is to measure data with the sources orthogonally
polarised, so the photons are entirely
distinguishable and no interference occurs
(\ref{equation2}).
\begin{subequations}
\begin{align}
\label{equation1}g^{(2)}_{\|}(0) =
\left(1+\frac{\eta}{\alpha^{2}}\right)^{-2}\\
\label{equation2}g^{(2)}_{\bot}(0) =
\left(1+\frac{2\eta}{\alpha^{2}}\right)\left(1+\frac{\eta}{\alpha^{2}}\right)^{-2}
\end{align}
\end{subequations}
Fig. \ref{Fig1} (b) plots $g^{(2)}_{\bot}(0)$ and
$g^{(2)}_{\|}(0)$ versus the ratio of source
intensities, $\eta/\alpha^{2}$. For
$g^{(2)}_{\|}(0)$, all coincidence counts at time
delay zero are due to multi-photon emission from
the laser, which falls as $\eta/\alpha^{2}$
increases. Using this simple analysis we can make
the surprising prediction that the visibility of
two-photon interference, $
(g^{(2)}_{\bot}(0)-g^{(2)}_{\|}(0))/g^{(2)}_{\bot}(0)$,
can approach unity as $\eta/\alpha^{2}$
increases.

\begin{figure}
\includegraphics[width=80mm]{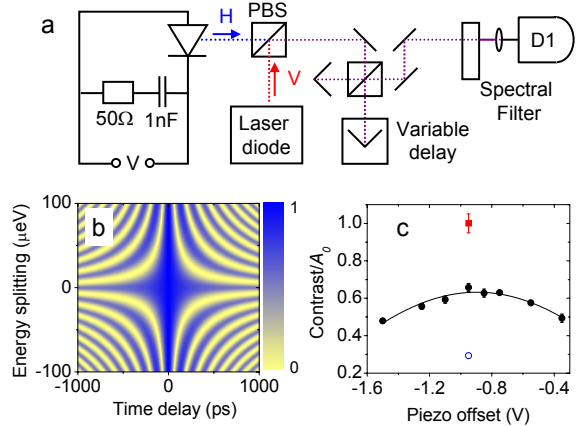}% Here is how to import EPS art
\caption{\label{Fig2} Measurements of
single-photon interference. (a) Layout for the
experiment (b) Predicted fringe contrast as a
function of the energy difference between the
sources and the delay in the Michelson
interferometer. (c) Experimental fringe contrast
for the dot and laser signals combined at 380
$ps$ delay as the bias applied to the
piezo-electric stack tunes the laser wavelength
(black). Also, shown are the fringe contrasts of
the laser (red) and dot (blue) at the same delay,
measured separately. Error bars represent
standard deviations determined from least-squares
fits to the data.}
\end{figure}

We now determine the coherence properties of the two sources used in our experiment. Single-photon
interference measurements were carried out using a free space Michelson interferometer with a
variable time delay \cite{Kammerer02} (Fig. \ref{Fig2} (a)). The interference pattern as a function
of delay is measured using an avalanche photodiode (D1). Looking at emission from the $X^{-}$ state
of the quantum dot source \cite{Bennett05} on its own we see that the interference has a fringe
contrast which varies as $A_{0}exp(-|t|/\tau_{coh})$ where $A_{0}$ is the fringe contrast at zero
delay, $t$ the delay time and $\tau_{coh}$ the coherence time. For the dot studied here
$\tau_{coh}=$ 285 $ps$ at 100$\mu A$. This characteristic exponential variation in contrast is an
indication that the state has a Lorentzian line-shape in energy of width 4.4$\mu eV$. Thus we can
be sure that homogeneous processes dominate the line broadening mechanisms \cite{Patel08}. We note
that the maximum fringe contrast observed at zero time delay, $A_{0}$, is below unity due to the
finite spatial overlap of light that travels along the two arms of the interferometer. Separately,
we have shown the coherence time of this source is sufficient to post-select interference events
between successive photons emitted by the source. The other source we employ is an external cavity
solid-state laser diode which can be tuned several hundred $\mu eV$ using a piezo-electric
actuator. This source has a coherence time of 1$\mu s$, which is three orders of magnitude longer
than we are able to probe with our Michelson interferometer. Thus, the fringe contrast is constant
at $A_{0}$ over the range of delays we probe.

In our experiment, to obtain a finite visibility
of two-photon interference these sources must
have the same energy to within the sum of their
linewidths \cite{Legero04}. However, our
spectrometer and CCD system only have a spectral
resolution $\sim$ 100$\mu eV$. Hence, to ensure
spectral indistinguishability we employ a scheme
based on single-photon interference, the layout
of which is shown in Fig. \ref{Fig2}(a). Clearly
the two photons have orthogonal polarisation at
the detectors and so will not give rise to
two-photon interference. However, the
single-photon interference patterns of the
separate sources have a period given by their
wavelength. Thus, when both sources are detected
at the same time we observe ``beating" in the
intensity at the detector. The period of the
beats is inversely proportional to the energy
difference between the states. We have developed
a simple model to illustrate how the
single-photon interference fringe contrast,
normalized to $A_{0}$, varies with both the
energy difference between the sources and
interferometer delay (Fig. \ref{Fig2} (b)). We
consider only the case where the sources appear
to have the same intensity on the detectors.

In practice, we set the interferometer delay to a
fixed value and measure the fringe contrast as a
function of the piezo-voltage applied to the
laser, which results in a near-linear variation
in laser energy. As can be seen in Fig.
\ref{Fig2}(c), the fringe contrast varies
cosinusoidally as a function of the energy
splitting between the two sources. For a delay of
380ps the period of the cosine variation is
34$\mu$eV. With a least squares fit to the
experimental data we can ensure degeneracy with
an error estimated below 1$\mu$eV, less than the
line-width of the broader source. Using this
method we have experimentally verified that the
sources' wavelengths remain stable within the
accuracy of this measurement over 24 hours.

\begin{figure}
\includegraphics[width=80mm]{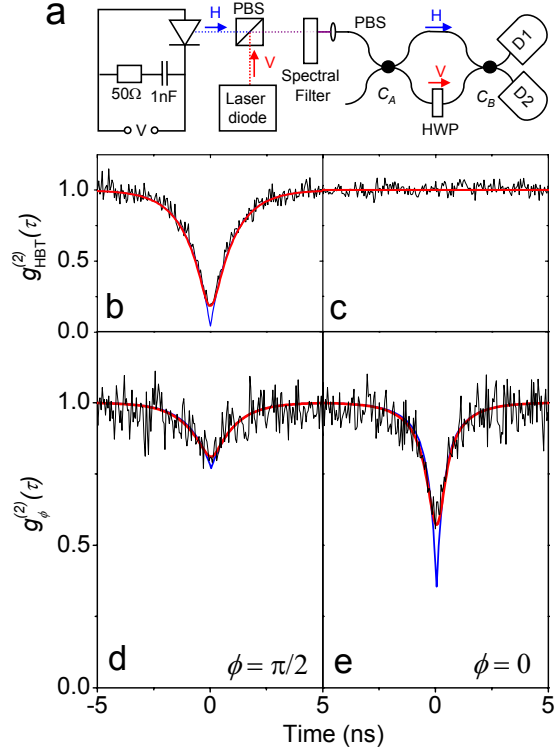}
\caption{\label{Fig3} Measurements of photon
statistics. (a) experimental layout for
two-photon interference between the sources.
Intensity correlation functions, recorded for (b)
the quantum light source only, (c) the laser
only, with both sources having (d) orthogonal and
(e) parallel polarisations. These plots show the
measured data (black), predicted correlations for
infinitely fast detectors (blue) and for the
measured detectors' response function (red).}
\end{figure}

We are now able to perform the two-photon
interference experiment using the apparatus in
Fig. \ref{Fig3}a. The co-linear and oppositely
polarised photons are passed to an interferometer
made of polarisation-maintaining optic fibre. The
first, polarising, coupler $C_{A}$ ensures every
photon from the dot takes the upper path to the
final non-polarising, $50/50$ coupler $C_{B}$ and
every photon from the laser takes the lower path.
This design increases the probability of the two
photons reaching the final coupler from opposite
sides by a factor of four, relative to previous
experiments \cite{Santori02}. Correlations at the
outputs of $C_{B}$ are measured with two silicon
avalanche photodiodes (APDs). A half-wave plate
in the path taken by the laser photon switches
its polarisation between being parallel and
orthogonal to the quantum dot's photon every few
minutes. This allows us to build up the
correlations for the case where the photons are
and are not interfering within the same
integration time. Thus any slow drift in the
position of the sources, or the fibres, which
might change the ratio of their intensities at
the detectors, is averaged out between pairs of
measurements. During the course of each
measurement the ratio of intensities is stable to
within 5 \%.

Fig. \ref{Fig3}(d) and (e) shows experimental
data recorded for equal intensity sources. The
measurement of $g^{(2)}_{\bot}(\tau)$ shows a dip
at time-zero due to the anti-bunched nature of
the quantum light source, as expected. More
strikingly, we can see a clear difference between
the measurements for parallel and orthogonal
polarisations, which is a result of two-photon
interference. This finite visibility constitutes
the main result of our experiment and occurs due
to interference between photons from the weak
laser and the anti-bunched source despite their
different linewidths and lack of common history.

To further quantify this result a full analysis,
including non-ideal source parameters, allows us
to calculate the correlation as a function of
time (equation \ref{EquationBig}).
\begin{equation}\label{EquationBig}
g^{(2)}_{\phi}(\tau)=R_{f}(\tau)\otimes\left[\frac{2\eta\alpha^{2}(1-\gamma^{2}\cos^{2}(\phi)\exp(\frac{-|\tau|}{\tau_{coh}}))+(\eta^{2}g^{(2)}_{HBT}(\tau)+\alpha^{4})}{(\eta +\alpha^{2})^{2}}\right]\\
\end{equation}
Where $\phi$ is the angle between the
polarisations of the two photons, $\gamma =
\langle\psi_{a}|\psi_{b}\rangle$ a measure of the
overlap of the two photon's wave-functions,
$R_{f}$ the detection system response function
and R = T. We note that in the case where
 $g^{(2)}_{HBT}(0) = 0$, $\gamma$=1 and $\tau = 0$ this
reverts to the form given in equations
\ref{equation1} and \ref{equation2}, as expected.
In this experiment $\eta$ and $\alpha^{2}$ are of
the order of $10^{-3}$. Separately we measure the
photon statistics of our sources using a
Hanbury-Brown and Twiss (HBT) arrangement
(equivalent to Fig. \ref{Fig3} (a) with only one
source operational at a time). For the laser the
auto-correlation function $g^{(2)}_{HBT}(\tau)
=$1 (Fig. \ref{Fig3} (c)) , as would be expected
for a photon source with Poissonian statistics.
For the QD source we expect anti-bunched emission
with a dip at time zero. We can predict the
precise shape of this auto-correlation
\cite{Patel08, Michler00} using the independently
measured radiative lifetime (985ps), the
contribution from background and dark counts
(which sum to 0.04 of the signal from the quantum
state) and the resolution of our detection system
(a Gaussian with width 428ps). This model
suggests $g^{(2)}_{HBT}(0)=0.19$, consistent with
our experimental measurement (Fig. \ref{Fig3}
(b)). From these parameters we calculate
$g^{(2)}_{\bot}(\tau)$ and $g^{(2)}_{\|}(\tau)$
for equal intensity sources being mixed. Shown as
blue lines in Fig. \ref{Fig3} (d) and (e) are the
correlations that would be observed for
infinitely fast detectors. However, when the
response function of the detection system is
included we obtain the curves shown in red. The
only free parameter is the wave-function overlap
$\gamma =$ 0.91.

\begin{figure}
\includegraphics[width=60mm]{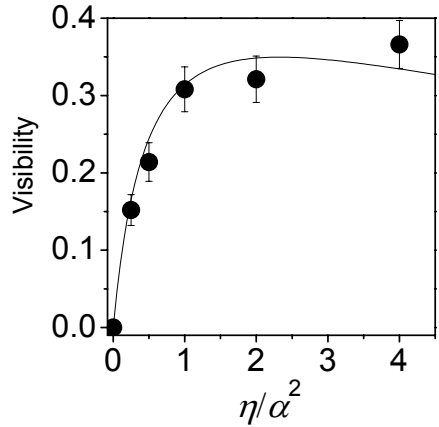}% Here is how to import EPS art
\caption{\label{Fig4} Measurement of the
visibility of two-photon interference as a
function of $\eta/\alpha^{2}$. Included as a
solid line is a fit using the known detector
response time and $\gamma = 91$ \%. Error bars
represent standard deviations determined from
least-squares fits to the data.}
\end{figure}

Finally, a series of measurements were made of
the visibility of interference as a function of
the intensity ratio of the two sources,
$\eta/\alpha^{2}$. Fig. \ref{Fig4} shows this
data and our prediction for $\gamma =$ 0.91. Our
theory predicts a maximum visibility will be
observed for $\eta/\alpha^{2}\sim 2$, which is a
result of the finite width of $R_{f}$. The
agreement between theory and experiment is good.
It is remarkable that we can infer such a high
overlap of the photons given the fundamental
differences in the sources. We note that the
maximum measured visibility of interference is
set by the ratio of the coherence time of the
solid state source to the response time of the
detection system. In future the raw visibility
could be increased by employing faster
superconducting detectors or long coherence time
atomic sources.

Acknowledgements. This work was partly supported
by the EU through the IST FP6 Integrated Project
Qubit Applications (QAP: contract number 015848).
EPSRC provided support for RBP and QIPIRC for
CAN.

Author contributions. RBP, CAN and DAR designed
and fabricated the samples. AJB and RBP carried
out the optical experiments. AJS guided the work.
AJB wrote the manuscript, with input from the
co-authors.

% *****     References     *****
%*******************************

\end{document}